\documentclass[letterpaper, 10pt, conference]{ieeeconf}      

\IEEEoverridecommandlockouts                              
\usepackage{amsmath}    
\usepackage{mathtools}    
\usepackage{amsfonts}
\usepackage{graphicx}   
\usepackage{subcaption}
\usepackage{epsfig} 
\usepackage{dsfont}
\usepackage{color}
\usepackage[normalem]{ulem}
\usepackage{cancel}
\usepackage{amssymb}
\usepackage{color}
\usepackage{bbm}
\usepackage{lipsum}
\usepackage[ruled,vlined,titlenotnumbered]{algorithm2e} 

\newcommand{\R}{\mathbb{R}} 
\newcommand{\veh}{Q} 
\newcommand{\cset}{\mathcal{U}}
\newcommand{\cfset}{\mathbb{U}}

\newcommand{\targetset}{\mathcal{L}}
\newcommand{\reachset}{\mathcal{V}}
\newcommand{\freachset}{\mathcal{F}}
\newcommand{\sfreachset}{\mathcal{S}}
\newcommand{\numfreachset}{F}
\newcommand{\dz}{\mathcal{Z}} 

\newcommand{\set}[1]{\{#1\}}            
\newcommand{\norm}[1]{\left\lVert#1\right\rVert}  
\newcommand{\abs}[1]{\left|#1\right|}  

\newcommand{\rom}[1]{\uppercase\expandafter{\romannumeral #1\relax}}

\usepackage{varioref}
\labelformat{algocf}{\textit{alg.}\,(#1)}

\newtheorem{defn}{Definition}

\newtheorem{thm}{Theorem}

\newtheorem{corr}{Corollary}

\title{\LARGE \bf
Predicting Stochastic Human Forward Reachable Sets Based on Learned Human Behavior}

\author{Jennifer C. Shih
\thanks{This research is supported by the NSF Frontiers project VeHICal, and by
the SRC CONIX program.}
\thanks{The author is with the Department of Electrical Engineering and Computer Sciences, University of
  California, Berkeley. cshih@berkeley.edu}
}

\begin{document}
\maketitle
\thispagestyle{empty}
\pagestyle{empty}

\begin{abstract}
  With the recent surge of interest in introducing autonomous vehicles to the
  everyday lives of people, developing accurate and generalizable algorithms for
  predicting human behavior becomes highly crucial. Moreover, many of these
  emerging applications occur in a safety-critical context,
  making it even more urgent to develop good prediction models for human-operated
  vehicles. This is fundamentally a challenging task as humans are often noisy 
  in their decision processes. Hamilton-Jacobi (HJ)
  reachability is a useful tool in control theory that provides safety
  guarantees for collision avoidance.
  In this paper, we first demonstrate how to incorporate information derived
  from HJ reachability into a machine learning problem which 
  predicts human behavior in a simulated collision avoidance
  context, and show that this yields a higher prediction accuracy than learning 
  without this information. Then we propose
  a framework to generate stochastic forward reachable sets
  that flexibly provides different safety probabilities and 
  generalizes to novel scenarios. We demonstrate that we can construct
  stochastic reachable sets that can capture the trajectories 
  with probability from $0.75$ to $1$.
\end{abstract}

\section{Introduction}

In recent years, there has been much excitement in introducing intelligent
systems that can navigate autonomously in environments with humans. For
example, many self-driving car companies \cite{Waymo}, \cite{Cruise},
\cite{Aurora} have emerged in the past few
years. Furthermore, projects like Amazon Prime Air \cite{Amazon16} and Google's Project
Wing \cite{GoogleDrone} aim to tap into the airspace for package delivery. 
There has also been immense interest in using UAVs for disaster response
\cite{AUVSI16}. While roads provide structure, in airspace the interaction of
vehicles occurs in a more unstructured setting, presenting the challenge of
making predictions based on more limited information.  

Critical to introducing autonomous vehicles into the workspace of humans is
safety. The authors in \cite{Mitchell05} have focused on safety from a differential
game perspective by characterizing the set of states from which
one vehicle is guaranteed to be safe from a second assuming 
the second can take worst case actions within a bounded set. In \cite{Akametalu2014}, 
the authors demonstrate learning these bounds online in an uncertain environment. 
The authors in \cite{Held17} characterize the notion of safety using torque 
limits on the robot. However, these works don't consider the
complexity of having humans in the environment. Taking humans into account is
challenging because unlike dynamics model governing the robotic systems, we
don't generally have models for how humans make decisions in unstructured scenarios.

Key to having safe human-robot interaction is the ability for robots to navigate
safely around human-operated vehicles. To enable this, the autonomous systems
need to make predictions of human behaviors to avoid collisions. Past work has used various 
methods to perform behavior prediction of vehicles operated by humans. Many
works have modeled humans as dynamical systems optimizing their own cost functions 
\cite{Abbeel04}, \cite{Ziebart08}. Inverse reinforcement learning aims to learn
these cost functions by observing past trajectories of the humans. There has
also been work on using recurrent neural networks to make predictions on future trajectories
\cite{Wu17}, \cite{Alahi16}. However, the methods presented in \cite{Abbeel04}-\cite{Alahi16} 
only generate a single trajectory and do not provide probabilistic information 
of future trajectories. Since human behavior is noisy,
having probabilistic information on future trajectories is important.

In the control theory literature, HJ reachability theory models the worst case
scenario \cite{Mitchell05} and hence is often overly conservative when 
applied to real world scenarios. There has been a growing body of work 
in using stochastic reachable sets to reduce conservatism and model
uncertainty. For example, \cite{Malone14} derives a stochastic reachable set and uses it
for motion planning by making assumptions on the behavior of moving obstacles
\textit{a priori}. However, humans often don't satisfy these assumptions and 
human behavior is also often influenced by the behavior of other vehicles
in the environment. 

In \cite{McPherson18}, the authors develop the notion of a human safe set for
a human supervisor, to model when the supervisor would 
start to intervene with robot teams operating on their own in order to
avoid static obstacles. However, this does not provide information on a mapping 
from \textit{any} joint configuration of the human and robot to the human's
action when we aim to also model how the human would avoid, for example,
predicting the direction of avoidance, and the entire avoidance process. In this paper, we also 
aim to model the situation in which humans are avoiding moving robots and comprehensively 
evaluate the effectiveness of incorporating varying
degrees of information derived from HJ reachability.

In \cite{Govindarajan17}, the authors learn a forward reachable set by 
optimizing the disturbance bound, learning from data
that is repeatedly gathered from similar initial configurations
and generating a reachable set that satisfies an accuracy threshold.
However, if we aim to apply the learned reachable set to a scenario
in which the two vehicles are approaching from an angle different from
what's seen in the training data set, the reachable set generated will not be suitable.
In \cite{DriggsCampbell18}, the authors assume that there are different modes that humans
are in and learn a reachable set for each of these distinct modes. In
this paper, we are interested in prediction methods that generalize to novel
scenarios without the need to train a reachable set for each scenario. 

In this paper, we first demonstrate how to incorporate
information derived from HJ reachability in learning human
behavior in a specific example of a simulated 
two vehicle unstructured setting. We illustrate that
the use of HJ reachability yields considerable improvement in predicting human
behavior when compared to not using it, for the humans
that participated in our study. Furthermore, we
propose a framework for learning stochastic human forward
reachable sets (SHFRS) that flexibly captures regions with varying
levels of safety. The proposed framework generalizes
to prediction in scenarios not trained on. We validate our approaches
on data gathered from human experiments with 8 participants.

The paper is organized as follows: section \ref{eq:background} presents the background of
HJ reachability in the context of two vehicle collision avoidance. Section
\ref{eq:method} presents our problem statement, how we incorporate information
derived from HJ reachability into the learning problem, and the stochastic
reachable set framework. Section \ref{eq:experiment} presents
our experiment setup, evaluation metrics, experimental results, and an implementation 
of the stochastic forward reachable set method described in \ref{eq:method}.
Section \ref{eq:discussion} includes conclusion and future work.  Due to space
constraints, all proofs are omitted, but can be found in the
full version of this paper on arXiv.


\section{Background \label{eq:background}}
Our proposed method builds on HJ reachability theory \cite{Mitchell05}. 
HJ reachability is a control-theoretic method that provides
safety certificates by characterizing the set of states that could,
under the worst case behavior of the unknown but bounded disturbance,
lead to danger.

%
%
We give a brief overview of how to apply HJ reachability to solve a pairwise
collision avoidance problem such as the one in \cite{Mitchell05}. Consider two
vehicles $\veh_1, \veh_2$ described by the following ordinary differential
equations (ODE):
\vspace{-0.5em}
\begin{equation}
\label{eq:vdyn} 
\dot{x}_i = f_i(x_i, u_i), \quad u_i \in \cset_i, \quad i = 1,2.
\end{equation}

Given the dynamics of the two vehicles \eqref{eq:vdyn}, we can derive the relative dynamics 
in the form of \eqref{eq:rdyn}: 
\vspace{-0.5em}
\begin{equation}
  \label{eq:rdyn} 
  \begin{aligned}
    \dot{x}_{ij} &= g_{ij}(x_{ij}, u_i, u_j) \\
    u_i &\in \cset_i, u_j \in \cset_j \quad i, j = 1, 2, \quad i\neq j.
  \end{aligned}
\end{equation}

We assume the functions $f_i$ and $g_{ij}$ are uniformly continuous, bounded, and Lipschitz continuous in arguments $x_i$ and $x_{ij}$ respectively for fixed $u_i$ and $(u_i, u_j)$ respectively. In addition, the control functions $u_i(\cdot)\in\cfset_i$ are drawn from the set of measurable functions\footnote{A function $f:X\to Y$ between two measurable spaces $(X,\Sigma_X)$ and $(Y,\Sigma_Y)$ is said to be measurable if the preimage of a measurable set in $Y$ is a measurable set in $X$, that is: $\forall V\in\Sigma_Y, f^{-1}(V)\in\Sigma_X$, with $\Sigma_X,\Sigma_Y$ $\sigma$-algebras on $X$,$Y$.}.

The set of states that represents a collision is denoted as
$\dz_{ij}$, and we compute the following backward reachable set
(BRS), which is the set of states from which a collision could
occur over $[0,t]$ based on the worst case action of $\veh_j$:
\vspace{-0.5em}
\begin{equation}
\label{eq:brs}
\begin{aligned}
&\reachset_{ij}(t) = \{x_{ij}: \forall u_i \in \cfset_i, \exists u_j \in \cfset_j, \\
&x_{ij}(\cdot) \text{ satisfies \eqref{eq:rdyn}}, \exists s \in [0, t],
x_{ij}(s) \in \dz_{ij}\}.
\end{aligned}
\end{equation}

Reachability theory is valid for any time horizon $t$; however, for clarity, 
we will let $t \rightarrow \infty$ in this paper. $\reachset_{ij}$ can be 
obtained as the sub-zero level set of the viscosity solution $V_{ij}(t, x)$ of a
terminal value HJ PDE. For details on obtaining $V_{ij}$, please see \cite{Mitchell05}. 
The BRS can thus be denoted as 
$\reachset_{ij} = \{x_{ij} \in \R^n: \lim_{t \rightarrow \infty} V_{ij}(t, x_{ij}) \le 0\}$. 
We will also use a slight abuse of notation and write $V_{ij}(x_{ij}) = \lim_{t
\rightarrow \infty} V_{ij}(t, x_{ij})$. The interpretation 
is that $\veh_i$ is guaranteed to be able to avoid collision with $\veh_j$ over an 
infinite time horizon as long as the optimal control is applied as soon 
as the potential conflict occurs, represented by the sub-zero level set of 
$V_{ij}(x_{ij})$.


\section{Methodology \label{eq:method}}
We aim to predict the behavior of a human controlling a vehicle (which
we will call $\veh_{\mathcal{H}}$ for human vehicle) in
a shared space with an autonomous vehicle (called $\veh_{\mathcal{R}}$ for robot
vehicle), in an unstructured setting. The unstructured setting presents the 
challenge of using a limited amount of information to correctly model humans. 
The methodology section consists of two parts. In the 
first, we frame the learning problem and propose a way to use information 
derived from HJ reachability in making better predictions. In the second, 
we propose a framework to generate stochastic human forward
reachable sets (SHFRS) using the behavior prediction obtained from the result of the learning problem
in the first part, although the proposed framework can work with any learning model
that outputs probabilistic prediction over actions of humans.
\subsection{Predicting human actions \label{sec:prediction}}


\subsubsection{Problem Statement}
In an environment with a human vehicle $\veh_{\mathcal{H}}$
and a robot vehicle $\veh_{\mathcal{R}}$ with dynamics in the 
form of \eqref{eq:vdyn}, the human operator controls $\veh_{\mathcal{H}}$ to 
avoid colliding with $\veh_{\mathcal{R}}$. 
The human can provide 
control from a discrete set of $M$ control inputs $\set{u_1, \dots, u_M}$ where
$u_{min} \leq u_m \leq u_{max}, m \in \set{1, \dots, M}$. For example, 
these could be one-dimensional real numbers corresponding to turning clockwise, turning
counter-clockwise, or going straight. 
%
Let the states of the human and robot be denoted as 
$x_{\mathcal{H}}$ and $x_{\mathcal{R}}$ respectively. 
%
We aim to learn to predict the human action 
$\hat{u}$ within $\set{u_1,\dots,u_M}$ at any joint configuration of $\veh_{\mathcal{H}}$ and $\veh_{\mathcal{R}}$. Our
first contribution is to show through a set of human experiments that
incorporating information derived from HJ reachability into the proposed learning problem
can improve accuracy in prediction.

\subsubsection{Proposed method}
In machine learning, feature
vectors are representations of data points that can
potentially improve the predictive performance. 
Since we're working in an unstructured environment, we avoid direct dependency on absolute states 
and use the relative state $x_{\mathcal{H}\mathcal{R}} = x_{\mathcal{H}} - 
x_{\mathcal{R}}$. 

The construction of good features is in general a challenging but important 
problem in machine learning \cite{Domingos12}. 
In this project, we choose from a set of "standard"
geometric features, and augment with features derived from the
safety value functions. The standard features we can incorporate 
are the translational and rotational components of the relative state. 
We also include cosine and sine of the rotational components in the 
relative states to provide nonlinear angular information. 
We denote features relevant to translation properties as $\vec{g}_t$, features relevant 
to rotational properties as $\vec{g}_r$, and features relevant to the trigonometric 
properties of the angles as $\vec{g}_{trig}$. For example, if the
state of the vehicle is described by 
$x_{\mathcal{H}} = \begin{bmatrix} x_{\mathcal{H}, 1} & x_{\mathcal{H}, 2} & x_{\mathcal{H}, 3}
\end{bmatrix}$ and  $x_{\mathcal{R}} = \begin{bmatrix} x_{\mathcal{R}, 1} &
  x_{\mathcal{R}, 2} & x_{\mathcal{R}, 3} \end{bmatrix}$ where the first,
second, and third components correspond to the $x$ coordinates, $y$ coordinates, and
rotational angles, the relative state is
$x_{\mathcal{H}\mathcal{R}} = \begin{bmatrix} x_{\mathcal{H}\mathcal{R}, 1}
  & x_{\mathcal{H}\mathcal{R}, 2} & x_{\mathcal{H}\mathcal{R}, 3} \end{bmatrix}
= \begin{bmatrix} x_{\mathcal{H},1} - x_{\mathcal{R},1} & x_{\mathcal{H},2} -
  x_{\mathcal{R},2} & x_{\mathcal{H},3} - x_{\mathcal{R},3} \end{bmatrix}.$ Then based on our
feature construction method, $\vec{g}_t \coloneqq
\begin{bmatrix} \abs{x_{\mathcal{H}\mathcal{R}, 1}} &
  \abs{x_{\mathcal{H}\mathcal{R}, 2}}
\end{bmatrix}^T$, $\vec{g}_r \coloneqq \begin{bmatrix}
  x_{\mathcal{H}\mathcal{R}, 3} \end{bmatrix}^T$, and $\vec{g}_{trig}
\coloneqq \begin{bmatrix} cos \left(x_{\mathcal{H}\mathcal{R}, 3}\right) & sin \left(
  x_{\mathcal{H}\mathcal{R}, 3} \right) \end{bmatrix}^T$. Combining
everything, we have that the standard feature vector has the form $
\vec{g}_{std} \coloneqq \begin{bmatrix} \vec{g}_t^T & \vec{g}_r^T & 
\vec{g}_{trig}^T  \end{bmatrix}^T.$

In safety critical scenarios, past works predominantly incorporate distance, 
measured by the l-2 norm, $\norm{\cdot}_2$, of
the translational properties in relative states, as a
feature in learning human behavior. For example, using the previous
example, the distance feature is $g_d \coloneqq \norm{\begin{bmatrix}
  x_{\mathcal{H}\mathcal{R}, 1} & x_{\mathcal{H}\mathcal{R}, 2}
\end{bmatrix}}_2$.

We hypothesize that safety levels derived from HJ reachability can potentially provide 
crucial information in predicting human action and the safety levels viewed from
different agents may both provide valuable information. 
Let the safety levels of the human with respect to the robot be denoted as
$V_{\mathcal{H}\mathcal{R}}(x_{\mathcal{H}\mathcal{R}})$ and the safety levels 
of the robot with respect to the human be denoted as $V_{\mathcal{R}\mathcal{H}}(x_{\mathcal{R}\mathcal{H}})$. 
If we include all the features proposed so far, we have the feature vector
  $\vec{g} = 
  \begin{bmatrix} 
    \vec{g}_{std}^T & 
    g_d & 
    V_{\mathcal{H}\mathcal{R}}(x_{\mathcal{H}\mathcal{R}}) &
    V_{\mathcal{R}\mathcal{H}}(x_{\mathcal{R}\mathcal{H}}) 
  \end{bmatrix}^T.$

Now we present how we learn the human actions given the construction of feature
vector $\vec{g}$. 
Let the data set be represented as $\set{x_{\mathcal{H}\mathcal{R},n},
u_n}_{n=1}^{N}$, where $N$ represents the number of data points,
$x_{\mathcal{H}\mathcal{R},n}$ represents the relative state of data point $n$,
and $u_n$ represents the action the human took at this relative state. 
We call $u_n$ the label for datapoint $x_{\mathcal{H}\mathcal{R},n}$. Let
$\vec{g}_n$ be the feature vector for $x_{\mathcal{H}\mathcal{R},n}$.

To model the problem probabilistically, we denote the
probability of label $u_n$ of a datapoint $x_{\mathcal{H}\mathcal{R},n}$
as the function $P(u_n | x_{\mathcal{H}\mathcal{R},n}; \theta)$ that is 
parameterized by $\theta$, with the goal to learn the parameter $\theta$.
We assume that 
the data points are independent and identically distributed (iid)
and we learn the parameter $\theta$ by maximizing the probability
\vspace{-0.5em}
\begin{equation}
  \underset{\theta}{max} \text{ } \prod_{n=1}^{N} P(u_n | x_{\mathcal{H}\mathcal{R},n}; \theta).
  \label{eq:mle}
\end{equation}
This is referred to as the maximum 
likelihood method. Here we refer to $P(u_n | x_{\mathcal{H}\mathcal{R},n};
\theta)$ as the function approximator. 

Alternatively, we could make no probabilistic assumption on the data.
For example, a Support Vector Machine (SVM) finds a hyperplane that separates 
data points with different labels. Another example is a decision tree, which
uses a tree-like structure to separate the data by recursively grouping it based
on the decision rule at each node of the tree.

\vspace{-0.5em}
\subsection{A framework for generating stochastic human forward reachable sets (SHFRS) based on
  learned human behavior \label{stoch_framework}}

\subsubsection{Problem statement}
We aim to provide a framework to generate stochastic
human forward reachable sets (SHFRS), with varying levels of safety, that capture future 
trajectories of $\veh_{H}$.
Mathematically, the problem is defined as follows: Let $T_{c}$ denote the
current time step. Given the past and current states of the trajectory, 
$\set{x_{\mathcal{H}}^{(i)}, x_{\mathcal{R}}^{(i)}}_{i=0}^{T_{c}}$, 
develop a framework to generate a SHFRS $\sfreachset$ 
composed of $\numfreachset$ mutually disjoint regions $\sfreachset_j$ such that
each region $\sfreachset_j$ is associated with a probability
$p_{\sfreachset_j}$.
Furthermore, the probabilities
$p_{\sfreachset_j}$'s should satisfy $\sum_{j=1}^{\numfreachset}
p_{\sfreachset_j} = 1.$

We say that a region provides a higher \textit{safety probability} if the
region captures future trajectories with a higher probability.
Our second main contribution is thus a framework that provides varying levels of 
safety probabilities through stochastic reachable sets in novel scenarios.

\subsubsection{Proposed method}
To generate a SHFRS satisfying the properties
described earlier, we learn $\numfreachset$ forward reachable
sets, $\freachset_1, \dots, \freachset_{\numfreachset}$ that satisfy the
following properties:
\vspace{-0.2em}
\begin{itemize}
  \item \label{property1} $\freachset_1 \subseteq \freachset_2 \subseteq \dots \subseteq
    \freachset_{\numfreachset-1} \subseteq \freachset_{\numfreachset}$. 
  \item \label{property2} We associate a probability $p_{\freachset_{j}}$ with each reachable set 
    $\freachset_j$. This probability specifies the likelihood that the trajectory in 
    the next $T$ time steps will be within $\freachset_{j}$.
    $p_{\freachset_1},\dots, p_{\freachset_{\numfreachset}}$ should satisfy
    $p_{\freachset_1} \geq
    \underline{p}$, $p_{\freachset_{\numfreachset}}=1$, and
    $p_{\freachset_{j+1}} \geq p_{\freachset_j}, \forall j \in \set{1, \dots,
    \numfreachset-1}$. Note that $\underline{p}$ represents the desired minimum
    probability with which
    the smallest reachable set $\freachset_1$ should capture the human
    trajectories.
\end{itemize}
\vspace{-0.2em}
Note that we define $\freachset_{0} = \emptyset$. With this definition
, $ p_{\freachset_0} = 0$. 
The set of constraints $p_{\freachset_{j+1}} \geq p_{\freachset_j}, \forall j
\in \set{1,\dots, \numfreachset-1}$ encourages the set of reachable sets to
provide increasing levels of safety. The constraint $p_{\freachset_{\numfreachset}}=1$ enables 
us to provide a safety certificate in the worst case scenario. 

We propose to generate these reachable sets $\freachset_j$'s by learning 
time-varying control input bounds for each.
For simplicity, we consider the case in which
the human input is one dimensional, though our method generalizes to the multi-dimensional case.

\begin{defn}
  \textbf{Time-varying bounds on control inputs:} Let $\underline{u}_{j}^{(i)}$ and
  $\boldmath \bar{u}_{j}^{(i)}$ denote the lower
  and upper bounds on the control inputs for reachable set $\freachset_j$,
  between time $t_{T_c+i}$ and $t_{T_c+i+1}$. This is defined for all 
  $\forall i \in \set{0, \dots, T-1}$, $\forall j \in \set{1, \dots,
  \numfreachset}$. Note that $t_{T_c+i}$ corresponds to the real-valued time
  at time step $T_c+i$.
\end{defn}

Given $\underline{u}_{j}^{(i)}$ and $\bar{u}_{j}^{(i)}$ for all $i \in \set{0,
\dots, T-1}$, we can construct the forward reachable set $\freachset_{j}$
using the following definition.

\begin{defn}
  \textbf{Forward reachable set with time-varying constraints on inputs}:
  Suppose we have time varying constraints for the control inputs, i.e., there exists $t_0, \dots, t_T$ such that
  $0 = t_0 < t_1 < \dots < t_{T-1} < t_T$ and the control input between time $t_i$ and
  time $t_{i+1}$ is within the set $U^{(i)} = [\underline{u}_{j}^{(i)},
  \bar{u}_{j}^{(i)}]$. Denote
  $\mathcal{L}$ as the initial set of states to grow the reachable set from. 
  To make this well defined on the
  boundary time points, we define the reachable set to be
  $\mathcal{F}(t_0) = \mathcal{L}$, $\mathcal{F}(t_{i+1}) = \set{x: 
  \exists u(t) \in U^{(i)} \text{ s.t. } t \in [t_{i},
  t_{i+1}], x(\cdot) \text{ satisfies } \dot{x} = f(x,u), x(t_i) \in \mathcal{F}(t_i), 
  x(t) = x}.$ 
  \footnote{In this definition, we adopt the convention used in hybrid systems, 
    in which the input bound switch corresponds to a mode switch, which allows us 
    to directly use the Level Set Toolbox \cite{LevelSet} from past work.}
\end{defn}

%
Algorithm 1 presents how we determine the time-varying bounds
$\underline{u}_{j}^{(i)}, \bar{u}_{j}^{(i)}$. The algorithm takes in the
following arguments: $\set{x^{(i)}}_{i=0}^{T_c}$, $G$,
$\set{\epsilon_j}_{j=1}^{\numfreachset}$, and $\set{k^{(i)}}_{i=0}^{T-1}$. 
We use $x^{(i)}$ as a shorthand notation for $(x_{\mathcal{H}}^{(i)}, x_{\mathcal{R}}^{(i)})$. 
For generality, the algorithm takes in
joint states from all past time steps and the current time step $T_c$. The
argument $G$ represents any learned function that can take in the past and current joint
configurations of the vehicles, and outputs a probabilistic distribution over 
the set of possible actions. Note that we allow $G$ to take in the past states
for prediction here, however, it could also
be the case that $G$ makes predictions solely based on the current joint
configuration like the function approximators learned in \ref{sec:prediction}. 
As we illustrate in the next paragraph as we
describe our Algorithm, the scalar $\epsilon_j$ determines how much to grow the
input bounds for $\freachset_j$, and the positive integer
$k^{(i)}$ indicates the number of likely actions we use to generate these
bounds at time step $i$.

Algorithm 1, which learns time varying bounds for the human inputs,
is described below. On line 1, "low" and
"high" predicted joint states are initialized with the values of 
the current joint states; here $\hat{x}_l^{(T_c+i)}, i>0,$ refers to 
the predicted joint states at time step $T_c+i$, using the low control input value, 
the subscript $h$ is used for the high control input value. 
Inside the loop, lines 3-4 obtain the $k^{(i)}$ top
likely actions based on the learned function $G$, for both the low 
and high cases; and lines 5-6 obtain
the corresponding lowest and highest value inputs, denoted
$\underline{u}^{(i)}, \bar{u}^{(i)}$ respectively. In lines 7-10, 
we obtain the corresponding desired $\underline{u}_j^{(i)}, 
\bar{u}_j^{(i)}$ for each reachable set $\freachset_{j}$ by expanding $\underline{u}^{(i)}, 
\bar{u}^{(i)}$ with $\epsilon_j$ as shown. Lines
11-12 obtain the robot's actions for both the $l$ and $h$ cases
using the function getRobotAction, which is
defined using the robot's model. On
line 13, the predicted actions of the human for the
case $l$ and $h$ at time step $T_c+i$ are set to be the lower and upper
bounds, respectively, of the control input for the 
smallest reachable set $\freachset_1$.
Lines 14-18 simulate the predicted states of the human
and robot for the next time step $T_c+i+1$. This process repeats
as $i$ increments. 

\begin{algorithm}
  \LinesNumbered
  \SetAlgoLined
  \SetKwInOut{Input}{input}
  \SetKwInOut{Output}{output}
  \Input{
    $\set{x}_{i=0}^{T_c}$,
    $G$,
    $\set{\epsilon_j}_{j=1}^{\numfreachset}$,
    $\set{k^{(i)}}_{i=0}^{T-1}$
  }
  \Output{
    $\underline{u}_{j}^{(i)}, 
    \bar{u}_{j}^{(i)}$ for $i \in \set{0, \dots, T_c-1}$, 
    $j \in \set{1, \dots, \numfreachset}$
  }

  Assign $\hat{x}_{l}^{(T_c)} = x^{(T_c)}$,
  $\hat{x}_{h}^{(T_c)} = x^{(T_c)}$\;
  \For{$i\gets 0$ \KwTo $T-1$}{
      $\text{set\_}\hat{u}_{\mathcal{H},l} $ =
      getTopKPredictedHumanActions(G, 
        $\set{x}_{i=0}^{T_c-1}$, 
        $\set{\hat{x}_{l}}_{i=T_c}^{T_c+i}$,
        $k^{(i)}$
      ) \;

      $\text{set\_}\hat{u}_{\mathcal{H},h}$ =
      getTopKPredictedHumanActions(G, 
        $\set{x}_{i=0}^{T_c-1}$, 
        $\set{\hat{x}_{h}}_{i=T_c}^{T_c+i}$,
        $k^{(i)}$
      ) \;

      $\underline{u}^{(i)} = \text{ getMin } \left( 
        \text{set\_}\hat{u}_{\mathcal{H},l},
        \text{set\_}\hat{u}_{\mathcal{H},h}\right)$ \;

      $\bar{u}^{(i)} = \text{ getMax } \left( 
        \text{set\_}\hat{u}_{\mathcal{H},l},
        \text{set\_}\hat{u}_{\mathcal{H},h}\right)$ \;

      \For{$j \gets 1$ \KwTo $\numfreachset$} {
        $\underline{u}_{j}^{(i)} = \operatorname{max}(\underline{u}^{(i)} - \epsilon_j, u_{min})$ \;
        $\bar{u}_{j}^{(i)} = \operatorname{min}(\bar{u}^{(i)} + \epsilon_j, u_{max})$ \;
      }

      $\hat{u}_{\mathcal{R},l}^{(T_c + i)} $ =
        getRobotAction(
        $\set{x}_{i=0}^{T_c-1}$, 
        $\set{\hat{x}_{l}}_{i=T_c}^{T_c+i}$
      ) \;

      $\hat{u}_{\mathcal{R},h}^{(T_c + i)} $ =
        getRobotAction(
        $\set{x}_{i=0}^{T_c-1}$, 
        $\set{\hat{x}_{h}}_{i=T_c}^{T_c+i}$
      )\;

      $\hat{u}_{\mathcal{H},l}^{(T_c + i)}, \hat{u}_{\mathcal{H},h}^{(T_c + i)}
      =\underline{u}_{1}^{(i)},  \bar{u}_{1}^{(i)}$ \;

      \For{boundType $\in \set{l, h}$}{
        \For{agentType $\in \set{\mathcal{H}, \mathcal{R}}$}{
          $\hat{x}_{\text{agentType,boundType}}^{(T_c+i+1)} = $ForwardDynamics( \
          $\hat{x}_{\text{agentType,boundType}}^{(T_c+i)}, 
          \hat{u}_{\text{agentType,boundType}}^{(i)})$ \;
        }
      }
  }
  \label{alg:reach_set_alg}
\caption{Generation of time-varying input bounds for reachable sets $\freachset_j$'s}
\end{algorithm}

We now present the conditions $\epsilon_j$'s should satisfy to enable 
$\freachset_{j} \subseteq \freachset_{j+1}, \forall j \in \set{1,\dots,\numfreachset-1}$.

\begin{thm}
\label{thm:reachset_property}
Based on this reachable set generation scheme, 
If for any $j \in \set{1, \dots, \numfreachset-1}$, $\epsilon_j \leq 
\epsilon_{j+1}$ and $\forall j \in \set{1, \dots, \numfreachset}, \epsilon_j
\geq 0$, for any $j \in
\set{1, \dots, \numfreachset-1}$, $\freachset_{j} \subseteq \freachset_{j+1}$.
\end{thm}

We compute $p_j$ associated with $\freachset_j$ by letting it be the percentage 
of the human trajectories that fall entirely within the predicted $\freachset_j$. 

\begin{corr}
  Using our algorithm, $p_{\freachset_{j+1}} \geq p_{\freachset_j}, \forall j \in
  \set{1,\dots,\numfreachset-1}.$
\end{corr}
\begin{corr}
  If we set $\epsilon_{\numfreachset}$ such that 
  $\underline{u}_{\numfreachset}^{(i)} = u_{min},
  \bar{u}_{\numfreachset}^{(i)} = u_{max}, \forall i \in \set{0,\dots,T-1}$, 
  then $p_{\freachset_{\numfreachset}} = 1$.
\end{corr}

Having $\sum_{j=1}^{\numfreachset} p_{\sfreachset_{j}} =
p_{\freachset_{\numfreachset}} = 1$ allows us to capture all possible future
trajectories and provides us with the ability to use 
$\freachset_{\numfreachset}$ as the most conservative safety standard.


\section{Experiments \label{eq:experiment}}
We conduct experiments to understand how incorporating information derived
from HJ reachability affects the prediction of human behavior with our proposed
method. Furthermore, we present the learned stochastic human forward 
reachable sets (SHFRS) generated based on our proposed framework with the 
same experimental data.

\subsection{Experimental design}
We recruited 8 participants between the age 20-27 from the university campus to
participate in the experiment. We gather trajectory data by having each human 
subject control a human vehicle in a simulated environment with another robot vehicle. 
The robot vehicle has a goal which is displayed to the human,
and the robot is automatically controlled to reach its
designated goal in exactly 10 seconds. The human subject is informed that
the robot is heading straight to the goal and will not actively avoid
the human vehicle. 

In our experiment, we use vehicles with the following dynamics: 
\begin{equation}
\label{eq:dyn_i}
\begin{aligned}
\dot{p}_{x,i} &= v \cos \psi_i \\
\dot{p}_{y,i} &= v \sin \psi_i \\
\dot{\psi}_i &= \omega_i, \quad \omega_{min} \le \omega_i \le \omega_{max}
\end{aligned}
\vspace{-0.5em}
\end{equation}

\noindent where the state variables $p_{x,i}, p_{x,i}, \psi_i$ represent the
$x$ position, $y$ position, and heading of vehicle $\veh_i$, $i \in
\set{\mathcal{H}, \mathcal{R}}$. Each vehicle travels at a constant speed 
of $v=2$, and its turn rate $\omega_i$ is constrained by $\omega_{min}=-0.5,
\omega_{max}=0.5$. 
Using a computer keyboard, the participants can provide control
inputs corresponding to $\set{\omega_{min}, 0, \omega_{max}}$, which are 
control inputs for turning right, going straight, and turning left respectively. 
We think that these controls are sufficient for the human participants to avoid the
robot vehicle while not burdening the cognitive load of the humans by giving
them an abundance of possible controls, which may confound the
experimental results.

\begin{figure}[h]
  \centering
  \begin{subfigure}[b]{0.22\textwidth}
    \includegraphics[width=\linewidth]{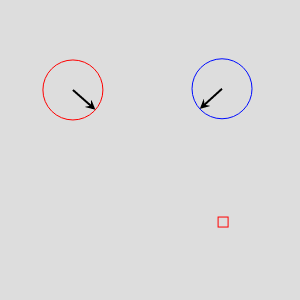}
  \end{subfigure}
  \begin{subfigure}[b]{0.22\textwidth}
    \includegraphics[width=\linewidth]{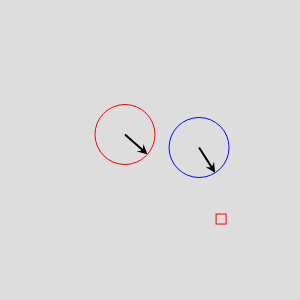}
  \end{subfigure}
  \\
  \vspace{0.1cm}
  \begin{subfigure}[b]{0.22\textwidth}
    \includegraphics[width=\textwidth]{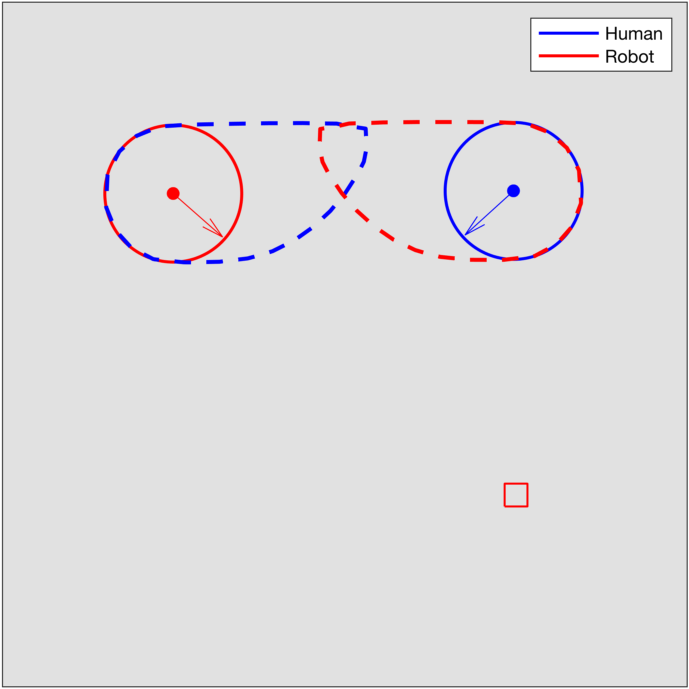}
  \end{subfigure}
  \begin{subfigure}[b]{0.22\textwidth}
    \includegraphics[width=\textwidth]{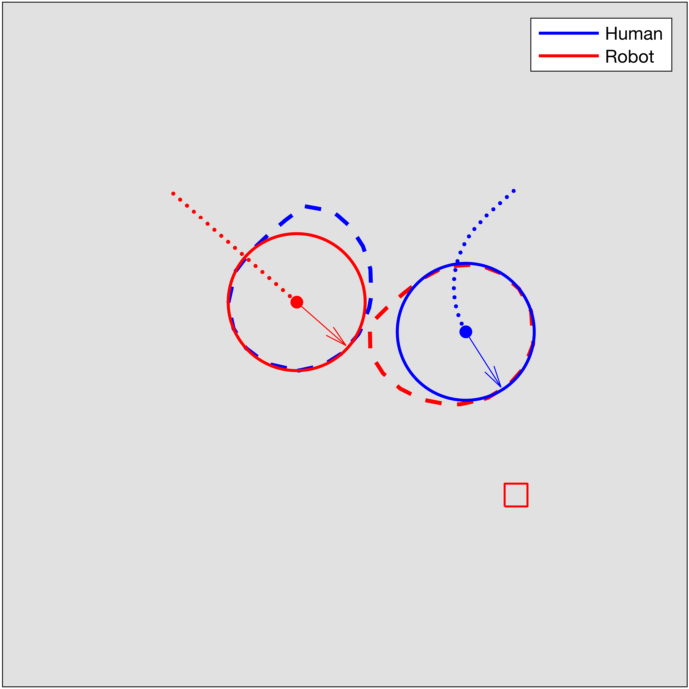}
  \end{subfigure}
  \caption{The top two figures illustrate the interface we designed to collect 
  data from participants. The top left figure shows the initial configuration
  and the top right figure shows the configuration after the human has inputted
  controls to avoid the robot. 
  The bottom figures illustrate the sub-zero
  level set for the value functions $V_{\mathcal{H}\mathcal{R}}$ and
  $V_{\mathcal{R}\mathcal{H}}$ for the
  configurations in the top figures, computed using \cite{LevelSet}. Neither of the
  safety value functions are provided to the human subjects
  during the experiment: the subjects only see the scenes in the
  top figures.}
  \label{fig:data_collection_scene}
  \vspace{-2.5em}
\end{figure}

Two example scenes from our data collection web application
are presented in the top two plots in Figure \ref{fig:data_collection_scene}. 
The robot and human vehicle are colored in red and blue respectively.
The tails and the directions of the arrows indicate the exact locations and
orientations of the vehicles respectively. The goal of the robot is presented as a red
square. The human operator is asked to provide control inputs so that the 
human vehicle avoids the danger zone of the robot vehicle,
represented as
$\targetset_{ij} = \{p_{x,ij}: (p_{x,i} - p_{x,j})^2 + (p_{y,i} - p_{y,j})^2 \le
R_c^2\}$, where $R_c=3$ in our experiment.
%
Each scene is also designed so that if the human
subject doesn't avoid the robot, the human and robot vehicles will
collide eventually. Hence the human subjects are instructed to continuously provide 
control inputs for avoidance starting from when they feel danger until they no longer think
that the human vehicle and the robot vehicle will collide if they don't provide
any more avoidance inputs. The human does not need to repeatedly press the key 
to avoid but can just hold down the key to avoid continuously. The scene ends when the robot reaches the goal.

The experiment is divided into three phases. In the first phase, we provide
instructions and the participants are given 1 minute 
to familiarize themselves with the interface we developed 
and the dynamics of the vehicle. In the second phase, participants are given three practice scenes
to further familiarize themselves with the setup of the study. In the
third phase, participants are given 50 vehicle avoidance tasks. The initial states 
of the human and robot vehicles are randomized for each scene. We record the
states of the human and robot vehicles, along with the action of the human, 
every 0.2 seconds. 

\vspace{-0.5em}

\subsection{Analysis of prediction performance \label{eq:experiment_pred}}
In this section, we present the experimental results of incorporating information
derived from HJ reachability into the learning problem.

%
We perform a five fold cross validation to tune the
hyperparameters of the models and evaluate the result based on 
predictive performance on the test data. We model 
that each human has a different pattern in avoidance behavior, hence, we 
trained a classifier for each subject. We perform an extensive comparison of
incorporating different subsets of information derived from HJ reachability. 
To rigorously compare the sets of features, we conduct statistical 
significance tests to see if the differences in performance 
are statistically significant. We
apply the Mann-Whitney test on pairs of feature sets and compute the p-values. 

As described in \ref{sec:prediction}, the standard features are $\mathcal{B} =
\set{\abs{p_{x,\mathcal{H}\mathcal{R}}}, 
\abs{p_{y,\mathcal{H}\mathcal{R}}}, \psi_{\mathcal{H}\mathcal{R}}, 
\cos \psi_{\mathcal{H}\mathcal{R}},
\sin \psi_{\mathcal{H}\mathcal{R}}}$. We augment these with features,
$d_{\mathcal{H}\mathcal{R}} = \sqrt{p_{x,\mathcal{H}\mathcal{R}}^2 +
p_{y,\mathcal{H}\mathcal{R}}^2}$, $v_{\mathcal{R}} =
V_{\mathcal{H}\mathcal{R}}(x_{\mathcal{H}\mathcal{R}})$, and 
$v_{\mathcal{H}} = V_{\mathcal{R}\mathcal{H}}(x_{\mathcal{R}\mathcal{H}})$ to
the feature set, which represent the distance between the two vehicles, 
and the safety levels of the human relative to robot and robot relative to
human, respectively. To obtain safety levels, we compute the BRS \eqref{eq:brs} 
with the relative
dynamics of the two vehicles derived from the vehicle dynamics \eqref{eq:dyn_i}.
To describe the sets of features, we use subscripts
$d$, $h$, and $r$ to represent the addition of the features to the standard
feature set
$d_{\mathcal{H}\mathcal{R}}$, $v_{\mathcal{R}}$,
and $v_{\mathcal{H}}$ respectively. For example, $\mathcal{B}_{hrd} =
\mathcal{B} \cup \set{v_{\mathcal{R}}, v_{\mathcal{H}},
d_{\mathcal{H}\mathcal{R}}}$ and $\mathcal{B}_{hr} = \mathcal{B} \cup
\set{v_{\mathcal{H}}, d_{\mathcal{H}\mathcal{R}}}$.

We conduct two sets of experiments: in experiment (\rom{1}), we 
perform prediction on the exact control the human inputs, i.e., we predict the
control as one of the three possible control inputs, $\set{\omega_{min}, 0,
\omega_{max}}$. In experiment (\rom{2}), the goal is to predict whether 
the human will input control to avoid at any joint state of the $\veh_{\mathcal{H}}$ 
and $\veh_{\mathcal{R}}$, which is equivalent to predicting whether the input is
$\omega=0$ or if the input falls in $\set{\omega_{min}, \omega_{max}}$ in our setup. 
We consider the following metric for both sets of experiments: 
\begin{itemize}
  \item Accuracy: $\frac{1}{\sum_{k=1}^{K} L_k}\sum_{k=1}^{K}\sum_{l=1}^{L_k} 
    \mathds{1}\{ \hat{y}_k^{(l)} = y_k^{(l)} \} \times 100\%$ where $\hat{y}_k^{(l)}$ and
    $y_k^{(l)}$ represent the predicted action and the ground truth action of
    the human at time step $l$ in trajectory $k$ respectively. Here $K$
    represents the number of trajectories and $L_k$ represents the number of time
    steps in trajectory $k$.
\end{itemize}

For experiment \rom{2}, we further evaluate the following metrics:
\begin{itemize}
  \item $D_{start}$: 
    Let $\hat{T}_{k,f}$ be the first time
    step in trajectory $k$ that our algorithm predicts the vehicle avoids and $T_{k,f}$ be
    the first time step the human avoids in the experiment. This metric is
    defined as: $\frac{1}{K}\sum_{k=1}^{K} \abs{\hat{T}_{k,f} - T_{k,f}}$.
  \item $D_{end}$: 
    Let $\hat{T}_{k,e}$ 
    be the last time step in trajectory $k$ that our algorithm predicts the
    vehicle avoids and $T_{k,e}$ be the last time step the human avoids in the
    experiment. This metric is
    defined as: $\frac{1}{K}\sum_{k=1}^{K} \abs{\hat{T}_{k,e} - T_{k,e}}$.
\end{itemize}

We consider support vector machine (SVM), decision tree (DT), and logistic
regression (LR) machine learning models. SVM and DT models don't make 
assumptions about the probabilistic distribution of the data. LR
assumes that each data point is iid. Despite the fact that the data we
gather are temporally correlated, we are interested in investigating how
well LR performs on the data.

Our experimental results indicate that incorporating the safety levels derived from
HJ reachability can yield considerable improvement in
predictive performance. Table
\ref{tab:accuracy} illustrates the accuracies obtained by applying the SVM,
DT, and LR models on the two sets of experiments, 
(\rom{1}) and (\rom{2}). We can see that for all models on both tasks, using the feature set
$\mathcal{B}_{hrd}$ yields the highest performance, considerably higher than
just incorporating distance, $\mathcal{B}_{d}$ ($p < 0.05$). It is also notable that for
all three algorithms on both tasks, incorporating exactly one of the two safety
levels $v_{\mathcal{H}}, v_{\mathcal{R}}$ outperforms incorporating just 
the distance $d_{\mathcal{H}\mathcal{R}}$ ($p < 0.05$). 
We see that 
incorporating the robot's safety level with respect to the human,
$v_{\mathcal{R}}$, generally, but not always, yields
higher performance than if not including it. This suggests that taking into
account the safety from the perspective of the robot vehicle is also important. 
With the feature set and model fixed, the accuracies in experiment (\rom{2}) is in 
general higher than those in experiment (\rom{1}). This is expected as in 
experiment (\rom{2}), we need to also predict the direction of
the avoidance. However, the difference is 
not big, suggesting that the algorithms did reasonably well in predicting
avoidance direction. 

Table \ref{tab:se_diff} illustrates the performance of the models 
on predicting the first and last time
steps of avoidance. 
We can see from the table that
using DT with feature set $\mathcal{B}_{hrd}$ yields the best
performance in metric $D_s$ and is on average $1.48$ time steps off from the 
ground truth, which is equivalent to $1.48 \times 0.2 = 0.296$
seconds off since every time step is $0.2$ second apart. 
Similar to the accuracy metric, 
incorporating just one of the safety levels yields a more accurate prediction
than just considering $d_{\mathcal{H}\mathcal{R}}$ ($p < 0.05$). We can also see
that including the feature $v_{\mathcal{R}}$ generally result in better
prediction than including the feature $v_{\mathcal{H}}$.
It is also interesting to see that the models did a much better
job predicting when a human would start avoiding than when a human would end
avoiding. We hypothesize that this is because humans are generally more noisy in
determining when to stop avoiding after there is no longer immediate danger as 
once the danger is cleared, when to stop avoiding is less important.


\vspace{-0.5em}
\begin{table}[h]
\begin{center}
  \begin{tabular}{ |c|c|c|c|c|c|c|c| } 
  \hline
  & $\mathcal{B}_{hrd}$ & $\mathcal{B}_{hr}$ & $\mathcal{B}_{hd}$ &
  $\mathcal{B}_{rd}$ & $\mathcal{B}_{h}$ & $\mathcal{B}_{r}$ &
  $\mathcal{B}_{d}$ \\
  \hline
  SVM (\rom{1}) & \textbf{78.67} & 78.46 & 76.56 & 76.74 & 75.11 & 76.33 & 69.75 \\
  \hline
  DT (\rom{1}) & \textbf{75.77} & 74.24 & 73.65 & 75.27 & 71.37 & 73.65 & 68.68 \\
  \hline
  LR (\rom{1}) & \textbf{77.73} & 77.4 & 74.71 & 77.38 & 73.31 & 76.61 & 69.15 \\
  \hline
  SVM (\rom{2}) & \textbf{83.89} & 83.15 & 81.12 & 82.85 & 79.62 & 82.15 & 70.72 \\
  \hline
  DT (\rom{2}) & \textbf{81.29} & 78.58 & 79.26 & 80.55 & 79.15 & 78.5 & 68 \\
  \hline
  LR (\rom{2}) & \textbf{81.70} & 81.61 & 78.79 & 81.62 & 77.54 & 80.88 & 71.35 \\
  \hline
  \end{tabular}
  \caption{This table shows the accuracies of SVM, DT, and LR models using different feature sets. We can see that including the information
derived from HJ reachability yields improvement in predictive performance
than just including distance as a feature.}
  \label{tab:accuracy}
\end{center}
  \vspace{-2.5em}
\end{table}

\begin{table}[h]
\begin{center}
  \begin{tabular}{ |c|c|c|c|c|c|c|c| } 
  \hline
  & $\mathcal{B}_{hrd}$ & $\mathcal{B}_{hr}$ & $\mathcal{B}_{hd}$ &
  $\mathcal{B}_{rd}$ & $\mathcal{B}_{h}$ & $\mathcal{B}_{r}$ &
  $\mathcal{B}_{d}$ \\
  \hline
  SVM, $D_{s}$ & 2.02 & 2.06 & 2.23 & \textbf{1.88} & 2.3 & 2.03 & 4.22 
  \\
  \hline
  DT, $D_{s}$ & \textbf{1.48} & 2.02 & 1.98 & 1.56 & 2.44 & 1.68 & 2.92 
  \\
  \hline
  LR, $D_{s}$  & 2.43 &  2.84 & 2.6 & \textbf{2.29} & 2.93 & 2.83 &  7.3
  \\
  \hline
  SVM, $D_{e}$ & \textbf{6.7} & 7.38 & 8.67 & 7.03 & 8.78 & 8.07 & 15.94 
  \\
  \hline
  DT, $D_{e}$ & 8.77 & 9.09 & 8.43 & \textbf{8.02} & 9.45 & 9.25 &  14.5 
  \\
  \hline
  LR, $D_{e}$ & 7.67 & \textbf{7.35} & 9.17 & 7.72 & 9.96 & 8.09 & 17.57 
  \\
  \hline
  \end{tabular}
  \caption{This table shows results for the metrics $D_{start}$ ($D_s$) and
  $D_{end}$ ($D_e$). Similarly to the accuracy metric, we see that both safety
  levels derived from HJ reachability improve prediction performance.
  }
  \label{tab:se_diff}
\end{center}
  \vspace{-2.0em}
\end{table}

Intuitively, we think the higher performance of including safety levels derived
from HJ reachability is that inherently the safety levels encode some information about
the dynamics and how dangerous the configuration is based on worst case
analysis. The geometric distance of the two vehicles
can also encode useful information about how dangerous the configuration might
be, however, it's not as informative as the safety levels.

\vspace{-0.6em}
\subsection{Stochastic human forward reachable set (SHFRS) implementation}
\begin{figure}[h]
  \centering
  \begin{subfigure}[b]{0.15\textwidth}
    \includegraphics[width=\textwidth]{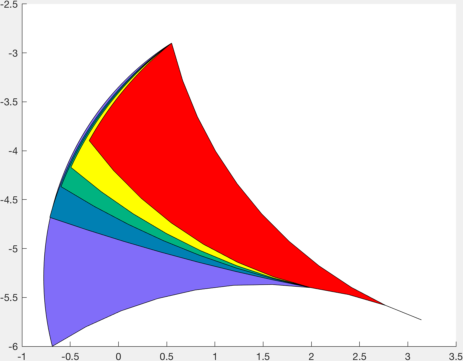}
  \end{subfigure}
  \begin{subfigure}[b]{0.15\textwidth}
    \includegraphics[width=\textwidth]{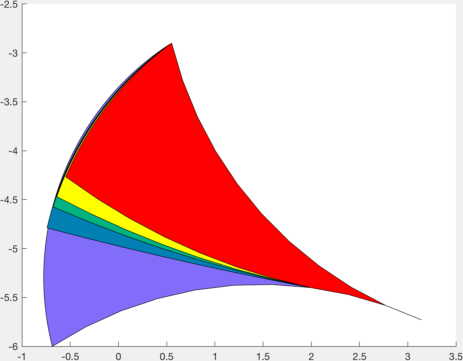}
  \end{subfigure}
  \begin{subfigure}[b]{0.15\textwidth}
    \includegraphics[width=\textwidth]{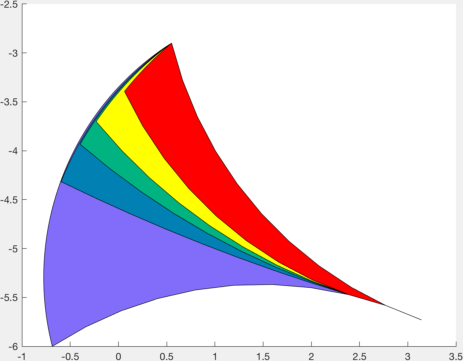}
  \end{subfigure}

  \caption{These are stochastic human forward reachable sets (SHFRS) generated using
  our framework. The algorithm predicts that the human will likely 
  turn right within the next $T=10$ time steps. 
  The regions in red, yellow, green, blue, purple correspond to $\sfreachset_1,
  \sfreachset_2, \sfreachset_3, \sfreachset_4, \sfreachset_5$ respectively.
  The middle figure is generated with the same $k^{(i)}$'s as the leftmost
  figure but the $\epsilon_j$'s used are larger or equal to those
  used in the leftmost figure. Increasing $\epsilon_j$ 
  increases the area of $\freachset_j$. The rightmost figure is generated with the same
  $\epsilon_j$'s as the leftmost figure, but with $k^{(i)}$'s smaller or equal 
  to those in the leftmost figure. We can see that decreasing $k^{(i)}$'s makes the
  areas of $\freachset_j$'s smaller.
    For the SHFRS in the middle figure, the probability for each of the five
    regions is:
    $p_{\freachset_1} = p_{\sfreachset_1} =0.752, 
    p_{\freachset_2} = p_{\sfreachset_1 \cup \sfreachset_2} = 0.771, 
    p_{\freachset_3} = p_{\cup_{i=1}^{3} \sfreachset_i} = 0.773,
    p_{\freachset_4} = p_{\cup_{i=1}^{4} \sfreachset_i} = 0.775, 
    p_{\freachset_5} = p_{\cup_{i=1}^{5} \sfreachset_i} = 1.0$.
  }
  \label{fig:stoch_reach_set_fix_k}
  \vspace{-1.0em}
\end{figure}

\vspace{-0.3em}
In this section, we demonstrate an implementation of our proposed forward
reachable set prediction framework in \ref{stoch_framework} by applying our 
framework on the experimental data gathered. 
For demonstration purpose, we use the logistic regression predictor learned in
experiment \rom{1} in \ref{eq:experiment_pred} for prediction, although 
it is possible to use \textit{any} predictor that gives probabilistic
information on the predicted actions. Note that the constraint $p_{\freachset_1}
\geq \underline{p}$ can always be satisfied by tuning $\epsilon_1$'s and $k^{(i)}$'s. 
The intuition is that the better the predictor we use in Algorithm I, the
smaller $\epsilon_1$ and $k^{(i)}$'s are needed to achieve $p_{\freachset_1}
\geq \underline{p}$. 
An example of the SHFRS using an implementation of this framework is illustrated
in Figure \ref{fig:stoch_reach_set_fix_k}.
We can see that the stochastic reachable set we produce gives varying degrees of
safety probabilities. 

To provide intuition on the effect of tuning $\epsilon_j$'s and
$k^{(i)}$'s during the optimization, we demonstrate the effect of changing these
parameters. The leftmost figure of Figure \ref{fig:stoch_reach_set_fix_k} shows the stochastic reachable set generated
with $\epsilon_1 = 0, \epsilon_2 = 0.15, \epsilon_3 = 0.25, \epsilon_4 = 0.4,
\epsilon_5 = 1.0$ and $k^{(i)} = 2, i \in \set{0,1}$ and $k^{(i)} = 1, i \in
\set{2, \dots, 9}$. The SHFRS in the middle figure is generated by fixing the $k^{(i)}$'s used in the left
figure and varying the $\epsilon_j$'s. We let $\epsilon_1 = 0.2$
$\epsilon_2 = 0.3, \epsilon_3 = 0.35, \epsilon_4 = 0.45$ and leave $\epsilon_5$
unchanged. Increasing $\epsilon_j$ 
makes the region $\freachset_j$ larger. On the other hand, the SHFRS in
rightmost figure is generated by
fixing the $\epsilon_j$'s used in generating the SHFRS in the leftmost figure
and varying the $k^{(i)}$'s. We let $k^{(i)} = 2, i = \set{0}$ and $k^{(i)} = 1,
i \in \set{1, \dots, 9}$ in the rightmost figure. Making $k^{(i)}$ smaller decreases the area of $\freachset_j$'s,
except for $\freachset_{\numfreachset}$, which aims to capture the worst case scenario.

%

Algorithm 1 can be computed efficiently online.
The generation of SHFRS based on the output of Algorithm 1 can be computed
online if the mapping from any $\set{[\underline{u}_{j}^{(i)},
\bar{u}_{j}^{(i)}]}_{i=0}^{T-1}$ to $\freachset_j$ has been computed offline.





\section{Conclusion and Future Work \label{eq:discussion}}
We first demonstrate how to incorporate information derived from HJ reachability
into a machine learning problem and show that this can yield considerable improvement
in prediction performance of human behavior 
compared with using one of the most typical features, the distance feature, in safety critical scenarios.
We then propose a framework to generate stochastic human forward reachable set
(SHFRS) that flexibly offers different levels of safety probabilities and generalizes to unseen
scenarios. In future work, it would be interesting to develop methodologies to evaluate 
how useful safety levels from HJ reachability are for continuous action prediction or under
situations where temporal information are considered. Other directions include further 
imposing metrics on the SHFRS framework and generalizing the framework 
to work with continuous action prediction. 


\section{Acknowledgement}
The author would like to thank Prof. Claire Tomlin for providing advice on the
project and helping to edit many parts of the paper.

\addtolength{\textheight}{1cm}   

 \bibliographystyle{IEEEtran}

\begin{thebibliography}{10}
\providecommand{\url}[1]{#1}
\csname url@samestyle\endcsname
\providecommand{\newblock}{\relax}
\providecommand{\bibinfo}[2]{#2}
\providecommand{\BIBentrySTDinterwordspacing}{\spaceskip=0pt\relax}
\providecommand{\BIBentryALTinterwordstretchfactor}{4}
\providecommand{\BIBentryALTinterwordspacing}{\spaceskip=\fontdimen2\font plus
\BIBentryALTinterwordstretchfactor\fontdimen3\font minus
  \fontdimen4\font\relax}
\providecommand{\BIBforeignlanguage}[2]{{%
\expandafter\ifx\csname l@#1\endcsname\relax
\typeout{** WARNING: IEEEtran.bst: No hyphenation pattern has been}%
\typeout{** loaded for the language `#1'. Using the pattern for}%
\typeout{** the default language instead.}%
\else
\language=\csname l@#1\endcsname
\fi
#2}}
\providecommand{\BIBdecl}{\relax}
\BIBdecl

\bibitem{Waymo}
\BIBentryALTinterwordspacing
{Waymo, Inc.} (2018) Waymo. [Online]. Available: \url{https://waymo.com/}
\BIBentrySTDinterwordspacing

\bibitem{Cruise}
\BIBentryALTinterwordspacing
{Cruise Automation, Inc.} (2018) Cruise. [Online]. Available:
  \url{https://getcruise.com/}
\BIBentrySTDinterwordspacing

\bibitem{Aurora}
\BIBentryALTinterwordspacing
{Aurora, Inc.} (2018) Aurora. [Online]. Available: \url{https://aurora.tech/}
\BIBentrySTDinterwordspacing

\bibitem{Amazon16}
\BIBentryALTinterwordspacing
{Amazon.com, Inc.} (2016) Amazon prime air. [Online]. Available:
  \url{http://www.amazon.com/b?node=8037720011}
\BIBentrySTDinterwordspacing

\bibitem{GoogleDrone}
\BIBentryALTinterwordspacing
{Business Insider}. (2017) Google's drone delivery project just shared some big
  news about its future. [Online]. Available:
  \url{https://www.businessinsider.com/project-wing-update-future-google-drone-delivery-project-2017-6}
\BIBentrySTDinterwordspacing

\bibitem{AUVSI16}
\BIBentryALTinterwordspacing
{AUVSI News}. (2016) Uas aid in south carolina tornado investigation. [Online].
  Available: \url{http://www.auvsi.org/blogs/auvsi-news/2016/01/29/tornado}
\BIBentrySTDinterwordspacing

\bibitem{Mitchell05}
I.~Mitchell, A.~Bayen, and C.~Tomlin, ``A time-dependent {Hamilton-Jacobi}
  formulation of reachable sets for continuous dynamic games,'' \emph{IEEE
  Transactions on Automatic Control}, vol.~50, no.~7, pp. 947--957, 2005.

\bibitem{Akametalu2014}
A.~K. Akametalu, J.~F. Fisac, J.~H. Gillula, S.~Kaynama, M.~N. Zeilinger, and
  C.~J. Tomlin, ``Reachability-based safe learning with gaussian processes,''
  in \emph{53rd IEEE Conference on Decision and Control}, Dec 2014, pp.
  1424--1431.

\bibitem{Held17}
D.~Held, Z.~McCarthy, M.~Zhang, F.~Shentu, and P.~Abbeel, ``Probabilistically
  safe policy transfer,'' in \emph{2017 {IEEE} International Conference on
  Robotics and Automation, {ICRA} 2017, Singapore, Singapore, May 29 - June 3,
  2017}, 2017, pp. 5798--5805.

\bibitem{Abbeel04}
P.~Abbeel and A.~Y. Ng, ``Apprenticeship learning via inverse reinforcement
  learning,'' in \emph{Proceedings of the Twenty-first International Conference
  on Machine Learning}, ser. ICML '04.\hskip 1em plus 0.5em minus 0.4em\relax
  New York, NY, USA: ACM, 2004.

\bibitem{Ziebart08}
\BIBentryALTinterwordspacing
B.~D. Ziebart, A.~L. Maas, J.~A. Bagnell, and A.~K. Dey, ``Maximum entropy
  inverse reinforcement learning,'' in \emph{Proceedings of the Twenty-Third
  {AAAI} Conference on Artificial Intelligence, {AAAI} 2008, Chicago, Illinois,
  USA, July 13-17, 2008}, 2008, pp. 1433--1438. [Online]. Available:
  \url{http://www.aaai.org/Library/AAAI/2008/aaai08-227.php}
\BIBentrySTDinterwordspacing

\bibitem{Wu17}
\BIBentryALTinterwordspacing
H.~Wu, Z.~Chen, W.~Sun, B.~Zheng, and W.~Wang, ``Modeling trajectories with
  recurrent neural networks,'' in \emph{Proceedings of the Twenty-Sixth
  International Joint Conference on Artificial Intelligence, {IJCAI-17}}, 2017,
  pp. 3083--3090. [Online]. Available:
  \url{https://doi.org/10.24963/ijcai.2017/430}
\BIBentrySTDinterwordspacing

\bibitem{Alahi16}
A.~Alahi, K.~Goel, V.~Ramanathan, A.~Robicquet, L.~Fei-Fei, and S.~Savarese,
  ``Social lstm: Human trajectory prediction in crowded spaces,'' in \emph{2016
  IEEE Conference on Computer Vision and Pattern Recognition (CVPR)}, June
  2016, pp. 961--971.

\bibitem{Malone14}
\BIBentryALTinterwordspacing
N.~Malone, K.~Lesser, M.~Oishi, and L.~Tapia, ``Stochastic reachability based
  motion planning for multiple moving obstacle avoidance,'' in
  \emph{Proceedings of the 17th International Conference on Hybrid Systems:
  Computation and Control}, ser. HSCC '14.\hskip 1em plus 0.5em minus
  0.4em\relax New York, NY, USA: ACM, 2014, pp. 51--60. [Online]. Available:
  \url{http://doi.acm.org/10.1145/2562059.2562127}
\BIBentrySTDinterwordspacing

\bibitem{McPherson18}
\BIBentryALTinterwordspacing
D.~L. McPherson, D.~R.~R. Scobee, J.~Menke, A.~Y. Yang, and S.~S. Sastry,
  ``Modeling supervisor safe sets for improving collaboration in human-robot
  teams,'' \emph{CoRR}, vol. abs/1805.03328, 2018. [Online]. Available:
  \url{http://arxiv.org/abs/1805.03328}
\BIBentrySTDinterwordspacing

\bibitem{Govindarajan17}
V.~Govindarajan, K.~Driggs-Campbell, and R.~Bajcsy, ``Data-driven reachability
  analysis for human-in-the-loop systems,'' in \emph{2017 IEEE 56th Annual
  Conference on Decision and Control (CDC)}, Dec 2017, pp. 2617--2622.

\bibitem{DriggsCampbell18}
\BIBentryALTinterwordspacing
K.~R. Driggs{-}Campbell, R.~Dong, and R.~Bajcsy, ``Robust, informative
  human-in-the-loop predictions via empirical reachable sets,'' \emph{{IEEE}
  Trans. Intelligent Vehicles}, vol.~3, no.~3, pp. 300--309, 2018. [Online].
  Available: \url{https://doi.org/10.1109/TIV.2018.2843125}
\BIBentrySTDinterwordspacing

\bibitem{Domingos12}
P.~Domingos, ``A few useful things to know about machine learning,''
  \emph{Commun. ACM}, vol.~55, no.~10, pp. 78--87, Oct. 2012.

\bibitem{LevelSet}
I.~M. Mitchell, ``{A Toolbox of Level Set Methods.}'' \emph{UBC Department of
  Computer Science Technical Report TR-2007-11 (June 2007)}.

\end{thebibliography}
  


\end{document}